\documentclass[aps,pre,twocolumn,floatfix,superscriptaddress]{revtex4-1}

\usepackage{amsmath}  
\usepackage{amsfonts} 
\usepackage{graphicx} 
\usepackage{epstopdf}
\usepackage{gensymb}
\usepackage{physics}
\usepackage{xcolor}
\usepackage{hyperref}
\usepackage{bm}

\begin{document}
\title{Physical rendering of synthetic spaces for topological sound transport}

\author{Hui Chen}
\thanks{These authors contributed equally to this work.}
\affiliation{Department of Mechanical and Aerospace Engineering, University of Missouri, Columbia, MO, 65211, USA}

\author{Hongkuan Zhang}
\thanks{These authors contributed equally to this work.}
\affiliation{School of Aerospace Engineering, Beijing Institute of Technology, Beijing 100081, China}

\author{Qian Wu}
\affiliation{Department of Mechanical and Aerospace Engineering, University of Missouri, Columbia, MO, 65211, USA}

\author{Yu Huang}
\affiliation{School of Aerospace Engineering, Beijing Institute of Technology, Beijing 100081, China}

\author{Huy Nguyen}
\affiliation{Department of Mechanical and Aerospace Engineering, University of Missouri, Columbia, MO, 65211, USA}

\author{Emil Prodan}
\email{prodan@yu.edu}
\affiliation{Department of Physics, Yeshiva University, New York, NY 10016, USA}

\author{Xiaoming Zhou}
\email{zhxming@bit.edu.cn}
\affiliation{School of Aerospace Engineering, Beijing Institute of Technology, Beijing 100081, China}

\author{Guoliang Huang}
\email{huangg@missouri.edu}
\affiliation{Department of Mechanical and Aerospace Engineering, University of Missouri, Columbia, MO, 65211, USA}

\begin{abstract}

Synthetic dimensions can be rendered in the physical space and this has been achieved with photonics and cold atomic gases, however, little to no work has been succeeded in acoustics because acoustic wave-guides cannot be weakly coupled in a continuous fashion. Here, we establish the theoretical principles and for the first time manufacture acoustic crystals composed of arrays of acoustic cavities strongly coupled through modulated channels to evidence one-dimensional (1D) and two-dimensional (2D) dynamic topological pumpings. In particular, the topological edge-bulk-edge and corner-bulk-corner transport are physically illustrated in finite-sized acoustic structures. We delineate the generated 2D and four-dimensional (4D) quantum Hall effects by calculating first and second Chern numbers and demonstrating robustness against the geometrical imperfections. Synthetic dimensions could provide a powerful way for acoustic topological wave steering and open up a new platform to explore higher-order topological matter in dimensions four and higher.

\end{abstract}

\maketitle

\section{Introduction}

%Topological matter is a rapidly growing field in which topological concepts are exploited to discover and classify new phases of matter \cite{Hasan2010,Qi2011,Chiu2016,Rachel2018}. {EP: the field is actually very mature and, for example, topological condensed matter peaked around 2013 and is cooling down now. Photonics also peaked.}

The physics of the integer quantum Hall effect (IQHE) \cite{KlitzingPRL1980} can manifest intrinsically in a condensed matter system \cite{HaldanePRL1988,ChangAdvMat2013,ChangScience2013} and same physics can be emulated with classical degrees of freedom, such as electro-magnetic \cite{Haldane2008,Wang2009,Rechtsman2013,hafezi2013imaging}, mechanical \cite{ProdanPRL2009,WangP2015,Nash2015,Wang_2015} and acoustic \cite{Fleury2014,Yang2015,Khanikaev2015,Fleury2016,Souslov2017}. Hall physics can be theoretically generalized to higher dimensions \cite{ZhangScience2001} and experimental realizations of the effect in 4D were recently reported with atomic and photonic systems as well as electric circuits \cite{LohseNature2018,zilberberg2017photonic,WangNaatComm2020}. These experimental works point to two different strategies for emulating higher space dimensions in physical space. The first approach \cite{WangNaatComm2020}
uses the higher dimensional lattice coordinates and orbital indices only as labels for resonators rendered in physical space, in order to connect them based on a standard higher dimensional model \cite{ProdanSpringer2016}. The second approach relies on synthetic dimensions generated via space modulations \cite{LohseNature2018,zilberberg2017photonic}, where the phases of the modulated structures can be used as adiabatic parameters that augment the physical space \cite{KLR2012}. In essence, these phase variation can be treated as introducing an additional global degree of freedom, usually called the phason, which lives on a torus. Even though the time-reversal symmetry is not broken for the static configurations of these systems, the IQHE physics emerges once the phason is pumped \cite{ProdanPRB2015}. As such, this second approach is extremely appealing because it does not require any active materials or other mechanisms to break the time-reversal symmetry.
%As noted in \cite{ApigoPRM2018}, no fine tuning is necessary to open topological gaps, provided the modulation and the couplings along the modulation are strong. As a result, these aperiodic principles are among the most practical methods to induce topological gaps.  

The synthetic dimensions generated by space modulations can be explored statically, \emph{i.e.} one point at a time \cite{ApigoPRM2018,ApigoPRL2019,NiCommPhys2019,VossPLA2020,hafezi2013imaging,xia2020topological,ChengArxiv2020}, which is usually achieved by manual reconfiguration of the system. The phason space can be also explored dynamically and here the most sought application is where the configuration of the system is modified cyclically in time such that the original dynamical pumping proposed by Thouless \cite{ThoulessPRB1983} can be observed. The challenge is to cycle fast enough to overcome the dissipation of a signal that self-oscillates as it is pumped from one edge to the other. This is very challenging and it has been only recently achieved experimentally for 1D but not for 2D pumping \cite{shen2019one,GrinbergNatComm2020,XiaArxiv2020,ChengPhysRevLett2020,XuXCPRL2020,NassarNRM2020}. A third strategy is to weakly couple modulated wave-guides, whose effective dynamics is described by a Schr\"odinger-like equation, with the coordinate along the wave-guides playing the role of time \cite{KLR2012}. This strategy has been experimentally implemented to produce topological pumping with photonic and elastic degrees of freedom \cite{KLR2012,verbin2015topological,RuzzenePRL2019,RivaPRB2020,zilberberg2017photonic,lustig2019photonic}.

The last strategy mentioned above is not feasible with sound because acoustic wave-channels cannot be weakly coupled in a continuous fashion along the guides and, to our knowledge, it has been never implemented with acoustic degrees of freedom. In this work, we present a distinct strategy that functions in the opposite regime, where the wave-guides are replaced by chains of coupled discrete resonators and strong couplings and modulations are also established in the transversal direction. In fact, the strategy we are proposing here can be better described as horizontal acoustic crystals carrying different phason values that are stacked and coupled with each other. By slowly varying the phason along the stacking direction, we demonstrate here that, with such approach, we can explore any continuous orbit inside the phason space, and even control the speed along this path to control the shape of the pumped pattern. As a result, we can render these abstract trajectories, occurring in the synthetic dimensions, on the physical dimension along the stackings. In turn, this enables us to control the propagation of the acoustic modes in space as well as the temporal phases of the signals.   

%The implications are extremely relevant for the cases where the phason space is complex, such as for 2D modulated crystals, where the phason lives on a torus. The later give access to the 4D IQHE physics and, as it is well known \cite{ProdanSpringer2016}, the hallmark feature of the topological boundary spectrum is a Weyl singularity \cite{ChengArxiv2020}, very much like the Dirac boundary singularity is for 3D topological insulators. To detect, demonstrate and take full advantage of the non-trivial topology of this singularity, one needs to explore phason trajectories that are not just straight lines, but also circles or loops. Our strategy will be even more relevant for the more elaborate phason engineerings that can generate even higher dimensional synthetic spaces, hence supplying access to higher IQHE physics \cite{ChengArxiv2020}.

With the unprecedented control over the phason, we demonstrate edge-to-edge topological pumping of sound in 1D modulated acoustic crystals, as well as edge-to-edge topological and corner-to-corner topological pumping in 2D modulated acoustic crystals. The higher-order topological corner-to-corner modes  in  our  system  are  principally  different  from conventional realizations. We demonstrate for the first time for these type of pumping processes that the topological sound transport is robust against random fluctuations in the resonator couplings. We also demonstrate that the pumping along a given orbit in the phason space occurs only in specific space directions. We delineate the generated 2D and 4D quantum Hall systems by calculating first and second Chern numbers. We also discuss various ways in which we can control these pumping processes and, moreover, we discuss new topological mode steerings in 2D modulated acoustic crystals that are entirely specific to the 4D IQHE physics.

 We believe that our work breaks ground for entirely new engineering applications, where the couplings in an acoustic crystal can be programmed for selective and robust point-to-point distribution of acoustic signals. We also theoretically predict that the phase of the sound signals can be also controlled with the same device. Compared with the traditional wave-guides, the virtual wave-guides proposed by our work can be reconfigured via small geometrical changes to the crystal. The latter, for example, can be implemented remotely, which will be highly desirable if the crystal is deployed in inaccessible places. There is practically no limit on how fast the reconfiguration of these virtual wave-guides can be. The principles demonstrated by our work can also be used to engineer new sensing devices, where these small geometrical changes occur in response to a stress, which then can be detected.

\section{Results}
\label{Sec:Results}

\subsection{Physical rendering of synthetic spaces}
\label{SubSec:PhysicalP}

We describe here the mechanism behind the physical rendering of a synthetic space and we start by describing the general setting. It consists of a generic acoustic structure of discrete resonators such that each resonator has an address $(\bm n,m)$, with $\bm n$ being a horizontal label and $m$ a vertical one. The couplings in the horizontal plane occur through channels whose modulated widths are specified by a phason $\bm \phi$, which lives on a phason space such as the $d_s$-torus. Here, $d_s$ represents the dimension of the synthetic space. The couplings in the vertical direction are established through uniform channels. Now, we consider any curve $\bm \phi(z)$ in the phason space with bounded first derivative $\bm \phi'(z)$, which we parametrize by the continuous $z$-coordinate. We use this curve to specify the phason values for each horizontal layer, specifically, $\bm \phi_m = \bm \phi(\epsilon z_m)$, where $z_m = m a_z$ is the physical coordinate of the $m$-th layer. The parameter $\epsilon$ is small and is there to ensure that the variations of the phason from one layer to another are small. Compared with the wave-guide setting, the major differences are the discrete character of the $z$ coordinate and the strong coupling in the horizontal direction.

The resulting dynamical matrix $D_{\bm \phi}$ governing the collective resonant modes depends on the chosen path $\bm \phi(z)$ and, while $D_{\bm \phi}$ is not periodic in the $z$-direction, it displays the following covariance property
\begin{equation}\label{Eq:Covariance}
T_v^\dagger D_{\bm \phi} T_v = D_{\bm \phi\circ \tau}, \quad \tau(z)=z+\epsilon a_z,
\end{equation}
where $T_v$ is the vertical translation by $a_z$. If $Q_{\bm n,m}$ is the acoustic resonant mode supported by the $(\bm n,m)$-cavity, then the pressure field of the resonant collective modes can always be sought in the form $\int d k_z \, \Psi_{k_z}(\bm r; \bm \phi)$ \footnote{See \emph{Supplementary Materials} for multi-mode expansions.}, with
\begin{equation}\label{Eq:GlobalMode}
\Psi_{k_z}(\bm r;\bm \phi) = \sum_{\bm n,m}e^{i k_z m}\varphi_{\bm n,m}(\bm \phi,k_z) Q_{\bm n,m}(\bm r),
\end{equation} 
and the covariance property~\eqref{Eq:Covariance} requires 
\begin{equation}
    \varphi_{\bm n,m+1}(\bm \phi,k_z) = \varphi_{\bm n,m}(\bm \phi\circ \tau,k_z).
\end{equation}
Since $\bm \phi(z)$ is a smooth path, then, $\bm \phi \circ \tau \approx \bm \phi + \epsilon a_z  \bm \phi'$ and, as such, $\varphi_{\bm n,m+1}(\bm \phi,k_z) \approx  \phi_{\bm n,m}(\bm \phi,k_z)$ to 0-th order in $\epsilon$. In these conditions, the vertical dispersion of the $\varphi$-coefficients can be ignored and the $\varphi_{\bm n,m}$ coefficients for a fixed $m$ become the eigen-modes of the reduced Hamiltonian (see \emph{Supplementary Materials})
\begin{align}\label{Eq:HEff}
H_{k_z}(\bm \phi_m)= & \sum_{\bm n}\nu(k_z)^2 \ket{\bm n}\bra{\bm n} \\ \nonumber 
& \qquad + \sum_{\langle \bm n,\bm n'\rangle} \kappa_{\bm n,\bm n'}(\bm \phi_m)\ket{\bm n}\bra{\bm n'},
\end{align}
where the last sum goes over the neighboring cavities and $\kappa_{\bm n,\bm n'}$ are the horizontal couplings, determined entirely by the phason value  $\bm \phi_m = \bm \phi( \epsilon z_m)$. Also, $\nu(k_z)$ is the dispersion of the decoupled vertical channels and, if $\epsilon(\phi_m)$ is the eigenvalue of the $\varphi$-mode, then the value of $k_z$ at layer $m$ is determined by the relation $f^2-\nu(k_z)^2=\epsilon(\phi_m)$. The conclusion is that, by examining the horizontal spatial profiles of the collective resonant modes, one layer at a time, we can visualize the states of the Hamiltonian $H(\bm \phi)$ along arbitrary paths inside the phason space.

\begin{figure*}
\centering
\includegraphics[width=.9\linewidth]{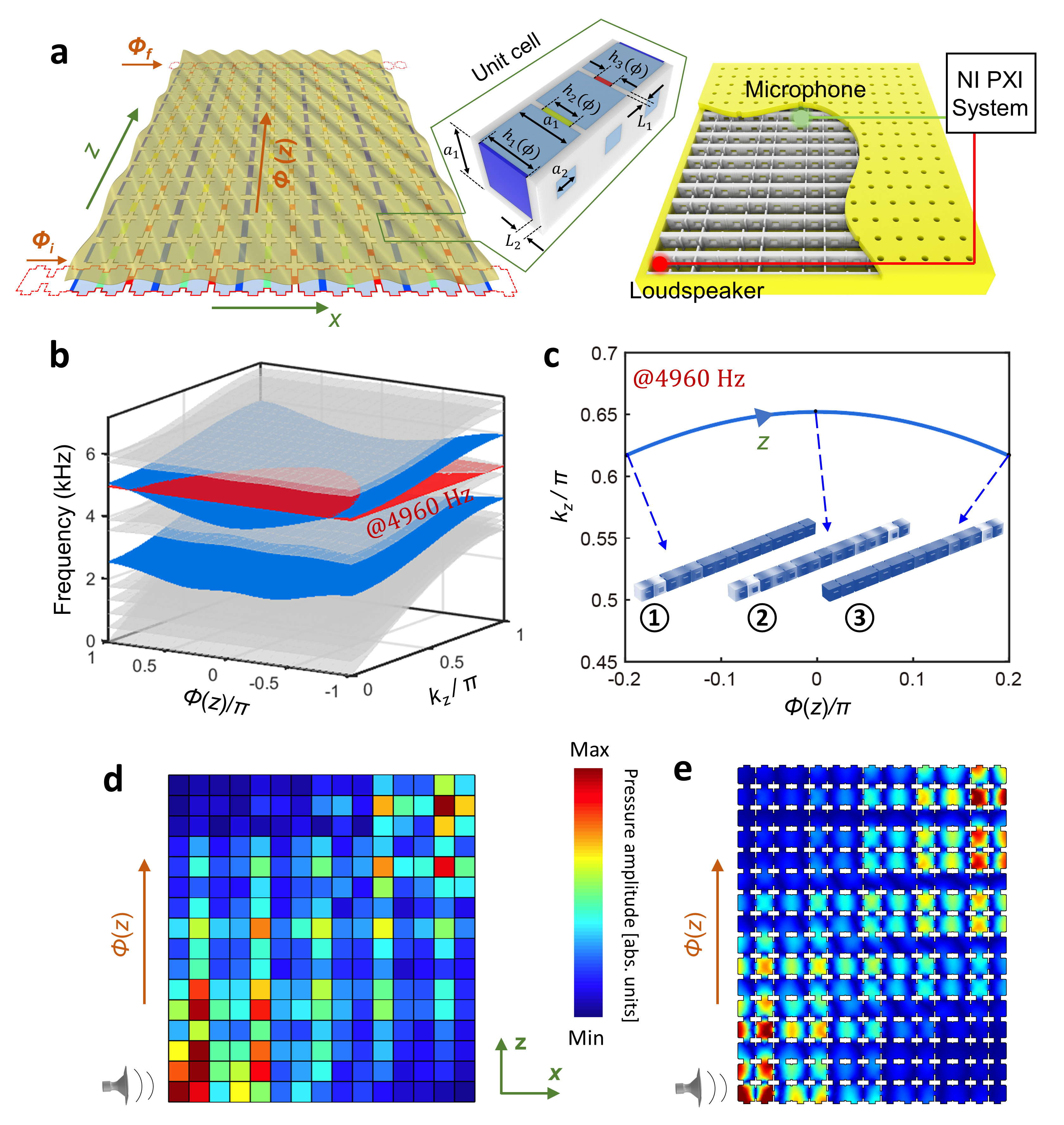}
\caption{{\bf 2D channel-modulated acoustic crystal with 1D topological pumping and its dispersion property.} ({\bf a}) Left panel: Schematic of the 2D modulated model of acoustic cavities coupled horizontally and vertically through channels. Inset: Geometry of the acoustic crystal consisting of coupled acoustic cavities connected with horizontal channels whose thickness $h_j=h_0[1+\delta \cos (b_j+\phi)]\ (j=1, 2, 3)$, where $h_0=12\ \mathrm{mm}$, $\delta = 0.6$, and $\{b_1,b_2,b_3\}=\{0,2\pi/3,4\pi/3\}$. The pumping parameter varies linearly from $\phi_i$ to $\phi_f$ along the $z$ direction. The side length of the cubic cavities is $a_1=20\ \mathrm{mm}$, the length of the modulated channels is $L_1=2\ \mathrm{mm}$, the length of the vertical connecting channels coupling the layers is $L_2=3\ \mathrm{mm}$, and their side length is $a_2=8\ \mathrm{mm}$. Right panel: Schematic of the printed 2D channel-modulated sample and the experimental setup. ({\bf b}) Dispersion diagram of a supercell composed of 15 coupled cavities, terminated by a hard-wall boundary along the $x$ direction, and a periodic boundary condition along the $z$ direction. The topological pumping of edge states is represented by blue surfaces, while the bulk bands are indicated by grey surfaces. ({\bf c}) Wave number $k_z$ as a function of the pumping parameter $\phi\in[-0.2\pi,0.2\pi]$ for frequency $f=4960\ \mathrm{Hz}$, corresponding to the cross section illustrated by the red dashed plane in (b). The insets show the topological mode from left to right localization with the change of the pumping parameter $\phi$ from $-0.2\pi$ to $0.2\pi$. Experimental ({\bf d}) and numerical ({\bf e}) demonstrations of topological edge pumping in the 2D system with a $15\times16$ array of coupled cavities (The excited frequency is $f=4960\ \mathrm{Hz}$). These results in (d) and (e) demonstrate that edge bands exist in the structure and appear on opposite sides of the device as a function of the pumping parameter, that is implied by the 2D Hall-type band structure of the system.}
\label{fig:fig1}
\end{figure*}

So far we have established that the coefficients $\varphi_{\bm n,m}$ for fixed $m$ are eigen-modes of Hamiltonian~\eqref{Eq:HEff}, but what is the weight and the phase of these modes in Eq.~\eqref{Eq:GlobalMode}? An expansion in the first order in $\epsilon$ (see \emph{Suplementary Materials}) reveals that the answer is supplied by the equation
\begin{equation}\label{Eq:Adiabatic}
    i \epsilon \Gamma(\phi) \partial_\phi |\varphi(\bm \phi) \rangle = [H(\bm \phi)-\epsilon(\bm \phi)]|\varphi(\bm \phi) \rangle, 
\end{equation}
which governs the evolution of the $\varphi$-coefficients along the $\bm \phi$-trajectory. The function $\Gamma(\bm \phi)$ is determined entirely by the vertical mode dispersion. After a proper change of variable, Eq.~\eqref{Eq:Adiabatic} becomes the classic equation of adiabatic evolution \cite{TeufelBook}, hence the amplitudes of the modes along the $z$ direction are all equal but a non-trivial phase $e^{i \alpha(z)}$ does develop along the stacking direction. It can be computed as the Berry phase of the Wilczek-Zee connection \cite{FrankPRL1984} along the path $\bm \phi(z)$. The important conclusion is that we not only can control the spatial profiles of the modes but also their phases. The latter have been already proposed as vehicles for certain forms of information processing with classical meta-materials \cite{BarlasPRL2020}.

\subsection{1D topological pumping}
\label{SubSec:1DPumping}

Figure~\ref{fig:fig1}a shows a planar array of acoustic cavities coupled horizontally and vertically through channels. Each cavity has an address $(n,m) \in \mathbb Z^2$ and the thickness of the horizontal channel connecting $(n,m)$ and $(n+1,m)$ resonators is modulated according to the protocol $h^x_{nm}=h_0[1+\delta \cos(b_{n\, {\rm mod}\, 3}+\phi_m)]$, where $h_0$ is the average thickness of the horizontal channels, $\delta$ is the modulation amplitude, and $b_j$'s are free parameters. The values of phason for each layer are set by $\phi(z) = \phi_i+(\phi_f-\phi_i)\frac{z}{L_z}$, where $\phi_i=-0.2\pi$, $\phi_f = 0.2\pi$ and $L_z=16a_z$. This results in a variation $\Delta \phi = 0.026 \pi$ from one layer to another, hence within the conditions of Sec.~\ref{SubSec:PhysicalP}.

By design, when $b_j=(j-1)\frac{2\pi}{3}$, the effective Hamiltonian $H_{k_z}(\phi)$ in Eq.~\eqref{Eq:HEff} is just the 1D Aubry-André-Harper (AAH) model \cite{Aubry1980} associated with the 2D Hofstadter model at magnetic flux $\pi/3$, with $\phi$ playing the role of a quasi-momentum. Figure~\ref{fig:fig1}b shows the resonant spectrum of $H_{k_z}(\phi)$ as function of $\phi$ and $k_z$. The computation was carried out with COMSOL Multiphysics on the domain of a finite horizontal stack with $k_z$-twisted Bloch boundary conditions in the vertical direction. The bulk and the boundary spectra are shown with distinctive colors and, as expected from the Hofstadter butterfly \cite{HofstadterPRB1976}, two bulk spectral gaps are observed. Also from the Hofstadter butterfly, one can read that the lower and upper band gaps carry first Chern numbers $C_1=\{-1,1\}$. This is confirmed in the \emph{Supplementary Materials} by direct calculations of the Chern numbers for an entire phase diagram computed for various values of $b_j$'s. Lastly, as expected from the bulk-boundary correspondence for the 2D IQHE, topological edge modes are observed in the spectrum reported in Fig.~\ref{fig:fig1}b (see the blue sheets). At fixed $k_z$, there are precisely one chiral edge band per edge and the slopes of these bands are consistent with the values of the Chern numbers (see \emph{Supplementary Materials}).

We now focus on the spatial profiles of the modes, as excited at frequency $f=4960\ \mathrm{Hz}$, indicated by the red horizontal sheet in Fig.~\ref{fig:fig1}b. It was chosen to intersect the dispersion surface of the edge modes such that we can visualize a topological pumping of sound. All modes along the curve resulted from the intersection of the $f=4960\ \mathrm{Hz}$ plane and the dispersion surfaces will be excited. As argued in the previous section, $\phi$ is resolved by the $z$ coordinate, hence this pumping curve, shown again in Fig.~\ref{fig:fig1}c, can be parametrized by the physical coordinate along the stacking, $(\phi(z),k_z(z))$. In other words, the spectral data from Fig.~\ref{fig:fig1}c has been rendered in the physical dimension, for us to observe. We now can understand the spatial profiles of the excited modes, when examined one stack a time. Along the horizontal stack at a given coordinate $z$, one should observe the mode of $H(\phi(z))$ at energy $f^2-\nu(k_z(z))^2$. The evolution of this mode along the pumping curve is shown in Fig.~\ref{fig:fig1}c, which confirms that sound is indeed pumped from one edge to the other. For example, the left edge state denoted in the inset (1) is selected as an initial state with a negative pumping value, which remains localized on the left boundary with the adiabatic increase of the pumping parameter and the corresponding wave number. When the pumping parameter approaches $\phi=0$, the left edge state becomes the bulk state as inset (2) depicted. As the pumping parameter increases further, the bulk state is then transformed into an edge state (3) localized at the right side. Then our prediction is that sound is transported from one side to the opposite side of the structure and this topological sound steering can be witness by walking along vertical coordinate. Experimental observation and confirmation of the adiabatic pumping via topologically protected boundary states is reported in Fig.~\ref{fig:fig1}d. Here, the sound was injected into the bottom-left corner of the structure and the pressure field was mapped by a microphone for each site (see the inset in Fig.~\ref{fig:fig1}(a)). As one can see, the pressure distribution indeed renders the topological pumping process in the physical dimensions. The experimental observation is also verified by the numerical simulation based on the exact geometry (Fig. \ref{fig:fig1}e, see also Supplementary Movie S1 for 1D transient topological edge pumping). The minor difference between experiment and simulation may be attributed to manufacturing deviations from connecting adiabaticity and perfect coupling to edge states (Details of the sample manufacturing, numerical simulation and experimental testing can be found in Methods). Let us also point out that there is substantial region where the wave has a bulk character and where dissipation mostly occurs. This region can be reduced by optimizing the function $\phi(z)$ from a linear to a tangent hyperbolic profile.

\begin{figure*}[b!]
\centering
\includegraphics[width=.9\linewidth]{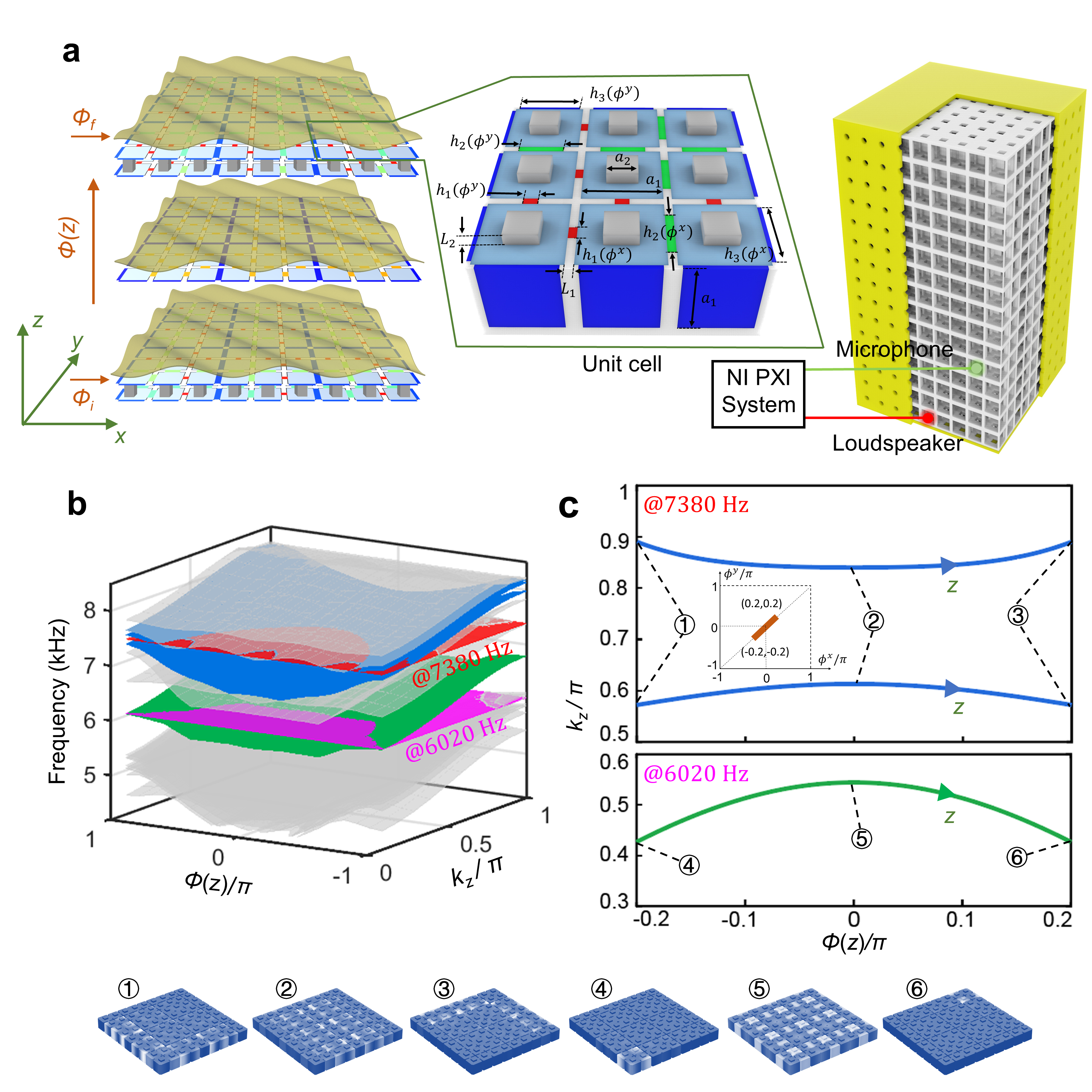}
\caption{{\bf 3D channel-modulated acoustic crystal with 2D topological pumping and its dispersion property.} ({\bf a}) Left panel: Schematic of the 3D channel-modulated acoustic crystal. Inset: Geometry of the acoustic crystal with horizontal connecting channels whose thickness $h_{\alpha j}=h_0[1+\delta \cos (b^{\alpha}_j+\phi^\alpha)]\ (\alpha\in\{x,y\}, j=1, 2, 3)$, where $h_0=11.2\ \mathrm{mm}$, $\delta = 0.75$, and $\{b_1^x=b_1^y,b_2^x=b_2^y,b_3^x=b_3^y\}=\{0,2\pi/3,4\pi/3\}$. The pumping parameters are varied linearly from $\phi_i$ to $\phi_f$ along the $z$ direction. The side length of the cubic cavities is $a_1=20\ \mathrm{mm}$ and the length of the modulated channels is $L_1=2\ \mathrm{mm}$. The length of the vertical connecting channels coupling the modulated layers is $L_2=3\ \mathrm{mm}$, and their side length is $a_2=8\ \mathrm{mm}$. Right panel: Schematic of the printed 3D channel-modulated sample and the experimental setup. ({\bf b}) Dispersion diagram of a supercell composed of $9\times9$ coupled cavities, terminated by a hard-wall boundary along $x$ and $y$ directions, and a periodic boundary condition along the $z$ direction. Bulk modes are shown in grey, topological pumping of edge modes in blue, and topological pumping of corner modes in green. ({\bf c}) Wave number $k_z$ as a function of the pumping parameter $\phi(z)\in[-0.2\pi,0.2\pi]$ for fixed frequencies $f=7380\ \mathrm{Hz}$ (upper panel) and $f=6020\ \mathrm{Hz}$ (lower panel), corresponding to the cross section illustrated by the red and purple dashed planes in (b). The inset in the upper panel shows the diagonal orbit (from $(-0.2\pi,-0.2\pi)$ to $(0.2\pi,0.2\pi)$) inside the phason space. The insets below (b) and (c) show representative mode shapes for each type of mode.}
\label{fig:fig2}
\end{figure*}

\subsection{2D topological pumping} 

We now investigate the three-dimensional (3D) acoustic structure shown in Fig.~\ref{fig:fig2}a, engineered to have a phason $\bm \phi=(\phi^x,\phi^y)$ living on 2-torus. In this case, each cavity has an address $(\bm n,m) \in \mathbb Z^3$ and the thicknesses of the horizontal connecting channels in the $\alpha=x,y$ directions are modulated according to the protocol $h^\alpha_{\bm n ,m}=h_0[1+\delta \cos(b^\alpha_{n_\alpha\, {\rm mod}\, 3}+\phi_m^\alpha)]$, while the vertical connecting channels are uniform. By design, the effective horizontal Hamiltonian~\eqref{Eq:HEff} is just a sum of two copies of the Hamiltonian from Sec.~\ref{SubSec:1DPumping}, $H(\bm \phi)= H(\phi^x) \otimes I + I \otimes H(\phi^y)$, which is known to host 4D QHE physics \cite{zilberberg2017photonic}.

The system can be pumped along any orbit inside the phason space by using the strategy described in Sec.~\ref{SubSec:PhysicalP}. The particular crystal in Fig.~\ref{fig:fig2}a was designed to pump along the diagonal orbit $\bm \phi(z) = (\phi(z),\phi(z))$, with $\phi(z)=\phi_i + (\phi_f -\phi_i)\frac{z}{L_z}$, where $L_z=15a_z$ (see the inset of Fig.~\ref{fig:fig2}c). In the \emph{Supplementary Materials}, we present additional 3D acoustic structures, engineered to pump along the orbit $(\phi(z),\phi_i)$. Figure~\ref{fig:fig2}b shows the dispersion surfaces of the resulting effective Hamiltonian~\eqref{Eq:HEff} as function of the pumping parameter $\phi$ and $k_z$. The computation, which was carried as for the 2D structure, reveals two bulk gaps and, in addition, two spectral sheets highlighted in blue for which the modes are localized along two edges, as well as one sheet highlighted in green for which the modes are localized at the corners. In the \emph{Supplementary Materials}, we demonstrate that both bulk gaps carry non-trivial second Chern numbers. Figure~\ref{fig:fig2}c (upper panel) shows the pumping curves resulted from the intersection of the dispersion diagram with the plane at frequency $f=7380\ \mathrm{Hz}$, with the latter highlighted in red in Fig.~\ref{fig:fig2}b. In contrast to the 1D topological pumping, there are more than one such pumping curve. Similarly, Fig.~\ref{fig:fig2}c (lower panel) shows the pumping curve resulted from the intersection with the plane at frequency $f=6020\ \mathrm{Hz}$, with the latter highlighted in purple in Fig.~\ref{fig:fig2}b. All these three pumping curves are parametrized by the $z$-coordinate and, as in the 1D case, we can predict that, if one examines the horizontal stack at coordinate $z$, one should observe the eigen-modes of $H(\bm \phi(z))$ at energy $f^2-\nu(k_z(z))^2$. Samples of these modes are rendered in the inset of Fig.~\ref{fig:fig2}c and, as one can see, all three pumping curves that we engineered are very special. Indeed, as one pumps along the blue contours, the mode is pumped from one pair of edges to the opposite pairs of edges, while if one pumps along the green curve the mode is pumped from one corner to the opposite corner. Both pumping processes proceed through a bulk delocalization transition.

%The insets illustrate the space evolution of the edge-bulk-edge and corner-bulk-corner states with the pumping parameters ($\phi(z)=\phi_x=\phi_y$) adiabatically scanning from $-0.2\pi$ to $0.2\pi$. The edge states map onto $(4-1)$-dimensional hypersurface states, while the second-order corner states map onto $(4-2)$-dimensional surface states in the 4D system. These states highlight the hypersurface/surface phenomena that are associated with the second Chern number \cite{Zilberberg2018} (Details of the numerical simulation are given in Methods).

\begin{figure}[ht]
\centering
\includegraphics[width=1\linewidth]{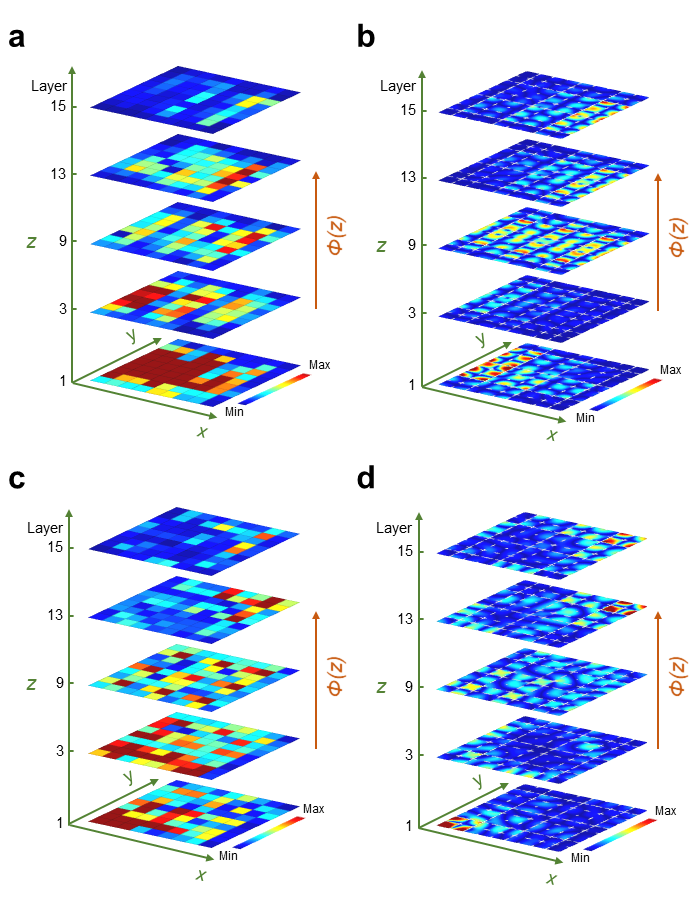}
\caption{{\bf Evolution of topological pumping of edge and corner modes in the 3D channel-modulated acoustic crystal with a $9\times9\times15$ array of coupled cavities.} Experimental {(\bf a)} and numerical {(\bf b)} observation of topological pumping of edge modes with the pumping parameters $\phi(z)$ scanning from $-0.2\pi$ to $0.2\pi$ (The excited frequency is $f=7380\ \mathrm{Hz}$). Acoustic waves are injected at the left-bottom edge to be pumped to the right-top edge. Experimental {(\bf c)} and numerical {(\bf d)} observation of topological pumping of corner modes with the pumping parameters $\phi(z)$ scanning from $-0.2\pi$ to $0.2\pi$ (The excited frequency is $f=6020\ \mathrm{Hz}$). Acoustic waves are injected at the left-bottom corner to be pumped to the right-top corner. These results demonstrate that edge and corner bands exist in the structure and appear on opposite sides of the device as a function of the pump parameters, that is implied by the 4D Hall-type band structure of the system.}
\label{fig:fig3}
\end{figure}

In Fig.~\ref{fig:fig3}a, we demonstrate experimentally that the predicted edge-bulk-edge pumping is indeed rendered in the physical space of the 3D structure. The acoustic wave is excited by a sound speaker along the bottom edge at $f=7380\ \mathrm{Hz}$ and the pressure distribution is measured by the microphone, layer by layer along the stacking direction (see the inset in Fig.~\ref{fig:fig2}a). As seen in Fig.~\ref{fig:fig3}a, the pressure field does evolve from the left to the right edge as one walks along the stacking direction. It is interesting to note that the edge states that we excite have the same energies as bulk states, which are also seen being excited in the very bottom layer. However, since the bulk states extend throughout the horizontal layer, they experience increased dissipation and, as such, they fade away along the stacking direction. On the other hand, the topologically pumped states are long-propagated, which further demonstrates the advantages of topological steering of sound. A numerical simulation was also conducted to validate our experimental observation (Fig. \ref{fig:fig3}b, see also Supplementary Movie S2 for 2D transient topological edge pumping). 

In Fig.~\ref{fig:fig3}c, we demonstrate experimentally that also the predicted corner-bulk-corner pumping can be rendered in the physical space of our 3D structure. Indeed, with source placed at a bottom corner and with the frequency adjusted at $f=6020\ \mathrm{Hz}$, one sees the measured pressure field evolving towards the opposite corner as one walks along the staking direction. This pumping is resolved much better than in the previous case, because the source couples less effectively to the bulk modes, hence the latter are not excited in this setup. Also, due to reduced dimensionality of the mode, the dissipation is weaker and the pressure field can be sustained longer along the stacking direction. Numerical simulations for the entire 3D structure are reported in Fig.~\ref{fig:fig3}c and they validate our experimental findings (see also Supplementary Movie S3 for 2D transient topological corner pumping).

%It is found that the left-corner modes in the bottom layer can be mostly injected to the bulk modes in the middle, and to the diagonally right-corner in the top layer. Such diagonal pumping under a concurrent scan ($\bm \phi(z) = (\phi(z),\phi(z))$) agrees with the 4D second Chern number bulk response. The acoustic diagonal pumping for both edge and corner modes through bulk bands is numerically expected (Fig. \ref{fig:fig2}c), in which each constituent 1D and 2D pumping is characterized by its own first and second Chern number, respectively (Details on the topological characterization are given in \emph{Supplementary Materials}). 

The emergence of the chiral boundary spectrum that makes possible the topological pumpings observed in Fig.~\ref{fig:fig3} is due to the second Chern number of the gaps, which is the strong topological invariant in 4D. As it is the case for any strong topological invariant, topological boundary spectrum emerges regardless of how the boundary of the crystal is cut, provided the available quasi-momenta are properly sampled. In our case, the phason plays the role of synthetic momenta and this implies that the pumping process along a given phason orbit manifests only in a particular space direction. To demonstrate that the pumping processes are indeed highly directional, we simulated the acoustic characteristics of the 3D structure for different phason orbits and with the source placed at different space locations and encountered the following scenarios: (1) propagation along a facet; (2) propagation along an edge; (3) pumping in the $x$ but not in the $y$ direction; (4) pumping in the $y$ but not in the $x$ direction; (5) pumping along the first diagonal $x=y$ but not along the second diagonal $x=-y$. (6) pumping along the second diagonal $x=-y$ but not along the first diagonal (see \emph{Supplementary Materials}). We also demonstrate what we call an ``antagonistic" effect, which manifests as follows: If edge-to-edge or corner-to-corner pumping is observed with the source placed on one edge or corner, respectively, then the pumping or any propagation are completely absent if the source is moved to the opposite edge or corner. In this respect, the structure acts like a perfect ``transistor" because the pumping can be turned on and off by $180^{\degree}$ rotations. In fact, by combining all the effects listed above, our 3D structure can be transformed into a multi-functional an programmable acoustic device for sound transport and distribution. We stress that our structure does not possess any crystalline symmetry so that the bulk polarizations are not quantized \cite{Petrides2020}. Therefore, the higher-order topological corner modes in our system are fundamentally different from previous realizations based on quantized quadrupole polarization \cite{Serra2018,Peterson2018,Imhof2018} or quantized Wannier centres \cite{NiX2019,Xue2019a,Xue2019b,Weiner2020}.

%or $y$ direction. In case (1), the system carries straight waveguides for either edge modes or corner modes along synthetic dimensions. In case (2), the acoustic diagonal pumping exists only when the pumping dimension is different from the dimension of the edge mode. Otherwise, either the edge or corner modes pumping is confined in the same pumping side without hybridizing with the bulk modes (see \emph{Supplementary Materials}).

\begin{figure}[ht!]
\centering
\includegraphics[width=1\linewidth]{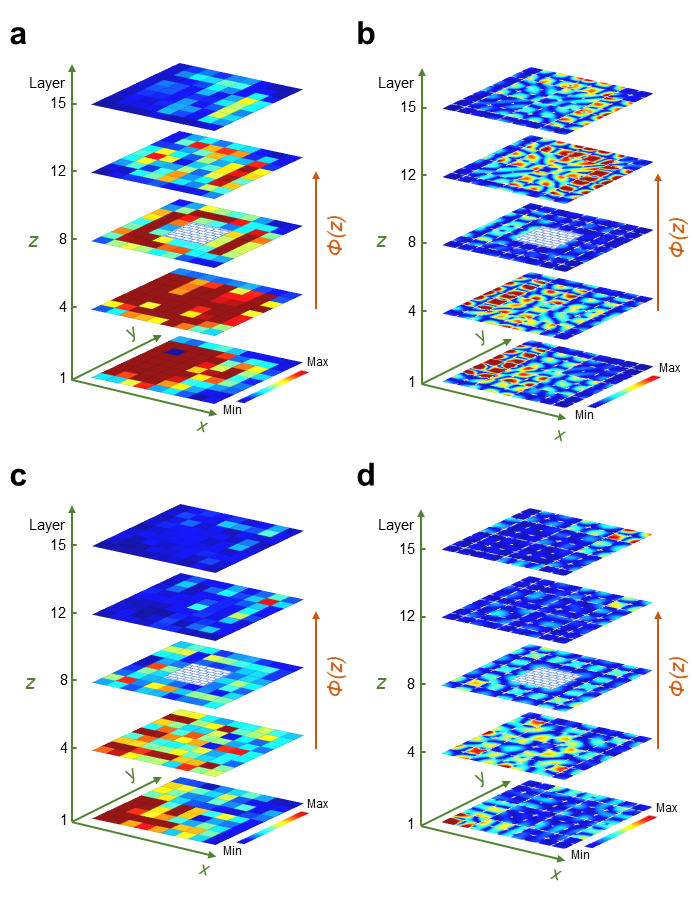}
\caption{{\bf Robust sound transport: evolution of topological pumping of edge and corner modes in the 3D channel-modulated acoustic crystal with a hollow.} Experimental {(\bf a)} and numerical {(\bf b)} observation of topological pumping of edge modes with the pumping parameter $\phi(z)$ scanning from $-0.2\pi$ to $0.2\pi$ (The excited frequency is $f=7380\ \mathrm{Hz}$). Acoustic waves are injected at the left-bottom edge to be pumped to the right-top edge. Experimental {(\bf c)} and numerical {(\bf d)} observation of topological pumping of corner modes with the pumping parameter $\phi(z)$ scanning from $-0.2\pi$ to $0.2\pi$ (The excited frequency is $f=6020\ \mathrm{Hz}$). Acoustic waves are injected at the left-bottom corner to be pumped to the right-top corner. The hollow area is indicated by the blue grid plane.}
\label{fig:fig4}
\end{figure}

Topological phases with non-zero second Chern numbers display intriguing wave transport characteristics, such as robustness against impurities or defects \cite{ProdanSpringer2016}. It is of interest to quantify the extent of the topological protection in such conditions. To evaluate that, a 3D hollow structure is constructed by removing nine cavities at the center of the system and the topological edge-bulk-edge and corner-bulk-corner pumpings in the defected structure are experimentally measured, as shown in Figs. \ref{fig:fig4}a and \ref{fig:fig4}c, respectively. When comparing performance with the system without defects (Figs. \ref{fig:fig3}a and \ref{fig:fig3}c), both edge and corner modes can be smoothly pumped despite of the presence of defects in a relatively large scale. This confirms that the topological pumping is immune against back reflections from defects or discontinuity. The experimental observation is also verified by the numerical simulation based on the exact geometry (Figs. \ref{fig:fig4}b and \ref{fig:fig4}d). In addition, a robust topological pumping due to disorder in the pumping parameters is also numerically evaluated in the \emph{Supplementary Materials}. 
%Finally, we note that the wave transport in the space-modulated structure is reciprocal but path-dependent, which is different from the time-modulated material. The acoustic pumping can function well along the synthetic dimension where topological edge/corner modes exist. However, no such pumping can be expected when sound is injected along the different paths where topological modes cannot be excited. The numerical simulation is conducted to confirm the path-dependent topological pumping (see \emph{Supplementary Materials}). Considering these remarkable properties, the acoustic metamaterials with synthetic dimension constitute a key step toward the practical implementation of robust waveguides based on topological pumping. 

\section{Discussion}

We demonstrated a novel and robust strategy to explore the global degrees of freedom of modulated wave media. While we have exemplified here only simple orbits inside the phason spaces, the method has no limitations on the geometry and topology of these orbits. For example, in the present study on the 2D pumping, we explored the fundamental loops $\mathcal C_x$ and $\mathcal C_y$ of the 2-torus (see the horizontal and vertical orbits in \emph{Supplementary Material}) as well as the diagonal orbit, which is topologically equivalent to the combination $C_x+C_y$. The 2-torus, however, supports an infinite number of topologically distinct paths, which can be in principle explored with the methods demonstrated in this work. It remains to be seen if the phases of the sound signals can be resolved as predicted in Sec.~\ref{SubSec:PhysicalP}, in which case the steering of the modes in both space and time domains could be controlled with the same device. In our opinion, this will open entirely new engineering applications in sensing and information processing. Also, by replacing the wave-guides with discrete coupled resonators, one now has the opportunity to engineer the dispersion with respect to $k_z$ quasi-momentum. This will involve modulations along the vertical direction and this opens a new dimension in the design space which is yet to be explored. 

In conclusion, we have evidenced the topological sound transport in modulated acoustic crystals through edge-to-edge topological and corner-to-corner topological pumpings associated with the 2D and 4D quantum Hall effects by physically rendering of synthetic spaces. These observations imply that the system is characterized by a non-zero Chern number and therefore the topological pumping is immune to bulk scattering and exhibits strong protection against design imperfections. The modulated acoustic crystals with synthetic spaces offers a new platform and route for efficient acoustic topological mode transport by engineering desired patterns on a phason-torus, and the higher-dimensional quantum Hall effect may provide unprecedented surface acoustic phenomena in the finite structure. The phason space augments the physical space and this opens a door to higher dimensional physics in acoustics and mechanics. Although we focused on the acoustic implementation using synthetic spaces, our approach can be generalized to other degrees of freedom, such as additional frequency dimensions can also be harnessed for the frequency modulation \cite{Schwartz2013,Bell2017}. Going forward, it will be important to develop and explore such broader connections, as the idea of topological matter in synthetic dimensions is very general and the extension of this approach to other complex orbits is much awaited.

\

\noindent{\bf Methods}

\begin{figure}[ht]
\centering
\includegraphics[width=1\linewidth]{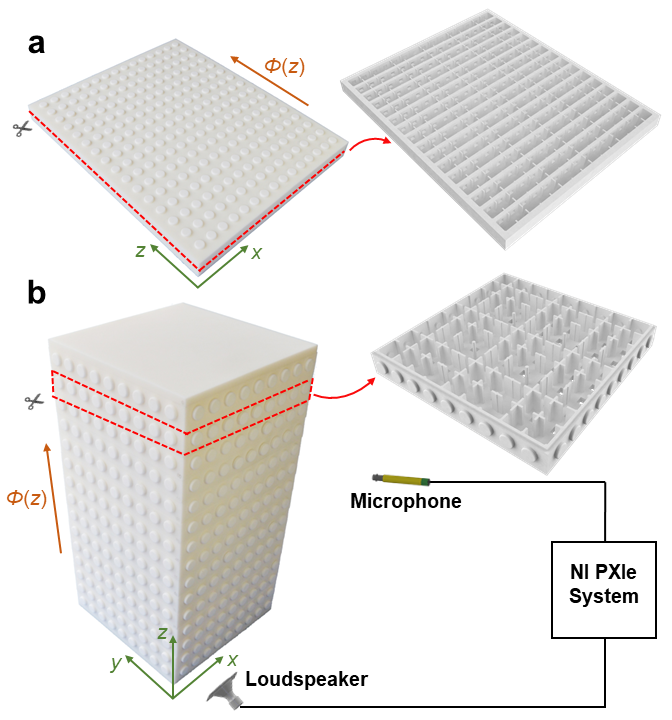}
\caption{{\bf Experimental samples and experimental setup.} {(\bf a)} Photograph of the printed 2D channel-modulated sample. {(\bf b)} Photograph of the printed 3D channel-modulated sample and schematic of the experimental setup.}
\label{fig:fig5}
\end{figure}

\noindent{\textit{Experimental specification} \textemdash} 
The stereolithographic (SLA) 3D printing technique is used to produce the experimental samples. The acoustic systems consisting of air cavities connected by modulated channels are made of photopolymer, which serves as acoustically hard walls due to a high impedance mismatch when compared with the air. The fabricated 2D channel-modulated sample (Fig. \ref{fig:fig5}a) consists of $15\times16$ cavities with thickness-modulated channels in the $x$ direction and thickness-constant channels in the $z$ direction. The fabricated 3D channel-modulated sample (Fig. \ref{fig:fig5}b) consists of $9\times9\times15$ cavities with thickness-modulated channels in the $xy$-plane and thickness-constant channels in the $z$ direction. The pumping parameters $\phi(z)$ in these two samples are evenly distributed from $-0.2\pi$ to $0.2\pi$ along the $z$ direction and other dimensional parameters are the same as the numerical model. 
Figure \ref{fig:fig5}b shows the experimental setup. A cylinder loudspeaker (diameter, $7\ \mathrm{mm}$; height, $5\ \mathrm{mm}$) and a microphone ($20\times 9\times 4\ \mathrm{mm}$) are used as the sound source and the acoustic pressure probe, respectively, both of which are small enough to be inserted in the cavities. The NI PXIe data acquisition system (PCI-4461 and PXIe-4610) is equipped for the experimental measurement, and a lab-made LabVIEW program controls the PCI-4461 to generate a sinusoidal signal that is amplified by PXIe-4610 to stimulate the speaker to emit sound waves of a given frequency. Meanwhile, the microphone is moved into different cavities to measure the sound pressure signals at the centre of each cavity and the data is recorded by the PCI-4461 data acquisition card. In each measurement, the unscanned holes are sealed by caps to avoid sound leakage. 

\noindent{\textit{Numerical simulations} \textemdash} 
The full-wave finite-element method simulations in this work are all performed using the commercial software COMSOL Multiphysics. The 3D geometry is implemented by filling with air (density $\rho=1.225\ \mathrm{kg/m^3}$ and speed of sound $v=343\ \mathrm{m/s}$). Eigenmode calculations within the ``Acoustic module'' are carried out to find the dispersion relations of the supercell. For calculations of the relation between the wave number and pumping parameters as well as the pressure eigenfunctions of the supercell, the simulations are implemented by the ``PDE (Partial Difference Equation) Interfaces module'' in which we write the coefficient form of the PDE with two independent variables: the density and the speed of air. Large-scale simulations are then implemented by the “Acoustic module” and frequency domain calculations are performed to obtain the steady acoustic pressure fields. 

\

\noindent{\bf Acknowledgements}
This work is supported by the Air Force Office of Scientific Research under Grant No. AF 9550-18-1-0342 and AF 9550-20-0279 with Program Manager Dr. Byung-Lip (Les) Lee, the NSF EFRI under Grant No. 1641078, the NSF CMMI under Award No. 1930873 with Program manager Dr. Nakhiah Goulbourne, the Army Research Office under Grant No. W911NF-18-1-0031 with Program Manager Dr. Daniel P. Cole. X.Z. acknowledges support from the National Natural Science Foundation of China (11872111, 11991030, 11991033, and 11622215) and 111 project (B16003). E.P. acknowledges financial support from the W.M. Keck Foundation and USA National Science Foundation through grant  DMR-1823800.

\

\noindent{\bf Author contributions}
H.C. and G.H. conceived the concept; H.C. and E.P. performed theoretical and numerical investigations; H.Z. and Y.H. conducted experiments; H.C., E.P. and G.H. analyzed methods and interpreted mechanisms, G.H. and X.Z. supervised the research; All the authors discussed the results and wrote the manuscript.

\

\noindent{\bf Competing interests}
The authors declare no competing interests.

\

\noindent{\bf Data and materials availability}
All data needed to evaluate the conclusions in the paper are present in the paper and/or the Supplementary Materials. Extended data, software, and materials in the main text and the Supplementary Materials are available upon request by contacting the corresponding authors.

\

\bibliographystyle{SciAdv}
\bibliography{Pumping}

\end{document}